\documentclass[english,nofootinbib,notitlepage,prl,
twocolumn]{revtex4}

\usepackage[T1]{fontenc}
\usepackage[latin9]{inputenc}
\usepackage{color}
\usepackage{graphicx}
\usepackage{amssymb,amsmath}
\usepackage{babel}

\def\be{\begin{equation}}
\def\ee{\end{equation}}

\makeatletter

\@ifundefined{definecolor}
 {\usepackage{color}}{}

\begin{document}

\title{ Standard Model False Vacuum Inflation:  Correlating \\ the Tensor-to-Scalar Ratio to the Top Quark and Higgs Boson Masses }

\author{Isabella Masina$^{1,2}$}
\email{masina@fe.infn.it}
\author{Alessio Notari$^{3}$}
\email{notari@ffn.ub.es}

\affiliation{$^{1}$  Dip.~di Fisica, Universit\`a di Ferrara and INFN Sez.~di Ferrara, Via Saragat 1, I-44100 Ferrara, Italy}
\affiliation{$^{2}$ CP$^3$-Origins \& DIAS, SDU 
Campusvej 55, DK-5230 Odense M, Denmark}
\affiliation{$^{3}$ Departament de F\'isica Fondamental i Institut de Ci\`encies del Cosmos, Universitat de Barcelona, Mart\'i i Franqu\`es 1, 08028 Barcelona, Spain}

\begin{abstract}
For a narrow band of values of the top quark and Higgs boson masses, the Standard Model Higgs potential develops 
a false minimum at energies of about $10^{16}$ GeV, where primordial Inflation could have started in a cold metastable state. 
A graceful exit to a radiation-dominated era is provided, {\it e.g.}, by scalar-tensor gravity models.
We pointed out that if Inflation happened in this false minimum, the Higgs boson mass has to be
in the range $126.0 \pm 3.5$ GeV, 
where ATLAS and CMS subsequently reported excesses of events.
Here we show that for these values of the Higgs boson mass,
the inflationary gravitational wave background has be discovered with a tensor-to-scalar ratio at hand of future experiments.
We suggest that combining cosmological observations with measurements of the top quark and Higgs boson masses
 represents a further test of the hypothesis that the Standard Model false minimum was the source of Inflation in the Universe.  
 \end{abstract}


 \maketitle



The fact that, for a narrow band of values of the top quark and Higgs boson masses, 
the Standard Model (SM) Higgs potential develops a local minimum \cite{CERN-TH-2683, hep-ph/0104016, strumia2} is nontrivial by itself, 
but it is even more suggestive that this happens at energy scales of about $10^{16}$ GeV, suitable for inflation in the early Universe. 

Inflation from a local minimum  is a viable scenario, provided a {\it graceful exit} to a radiation-dominated era can be obtained. 
Developing an explicit model with graceful exit in the framework of a scalar-tensor theory of gravity \cite{hep-ph/0511207,astro-ph/0511396}, 
in ref.\cite{Masina:2011aa} we pointed out that 
the hypothesis that inflation took place in the SM false vacuum is consistent only with a narrow range of values of the Higgs boson mass,
$m_H=(126.0 \pm 3.5)$ GeV, the error being mainly due to the theoretical uncertainty of the 2-loops Renormalization Group Equations (RGE) 
used in the calculation. 
This mass range is surprisingly compatible with the window $124-127$ GeV,
were both ATLAS and CMS  \cite{HCP11} recently reported excesses of events, in the di-photon as well as the 4-lepton Higgs decay channels. It is also compatible with preliminary results from Tevatron \cite{TEV}. 

These preliminary but very suggestive results provide a strong motivation to further 
investigate the scenario of SM false vacuum inflation, in particular by looking for complementary experimental tests.
Inflation can generate tensor (gravity wave) modes as well as scalar (density perturbation) modes. 
It is most common to define the tensor contribution through $r$, the ratio of tensor-to-scalar perturbation spectra at large scales. 
If inflation happened at a very high scale, as is the case for the SM false vacuum scenario,  quantum fluctuations during inflation
produced a background of gravitational waves with a relatively large amplitude.

In this Letter we argue that the tensor-to-scalar ratio, combined with the top quark and Higgs boson mass measurements,  
does represent a test of the hypothesis that inflation started from the SM false vacuum.


Let us consider the Higgs potential in the SM of particle physics.
For very large values of the Higgs field $\chi$, the quadratic term $m^2 \chi^2$ can be neglected and we are left with the quartic term, 
whose dimensionless coupling $\lambda$ depends on the energy scale, which can be identified with the field $\chi$ itself:
\begin{equation}
V(\chi) \simeq \lambda(\chi) \, \chi^4\,\,.
\end{equation}
It is well known that, for some narrow band of the Higgs and top masses, the Higgs potential develops a new local minimum 
\cite{CERN-TH-2683, hep-ph/0104016, strumia2}. 

If the Higgs field is trapped in a cold coherent state in the false minimum $\chi_0$ and dominates the energy density of the Universe, 
the standard Friedmann equation leads to a stage of inflationary expansion 
\be
H^2  \simeq \frac{V(\chi_0)}{3 M^2} \equiv H_I^2 \,\,\,,\,\,\,  \,\,a(t)\propto e^{H_I t} 
\label{eq-M}
\ee
where  $a(t)$ is the scale factor, $H \equiv \dot a /a$ is the Hubble rate and $M$ is the Planck mass.

A nontrivial model-dependent ingredient is how to achieve a graceful exit from inflation, that is a transition to a radiation-dominated era, in a nearly flat Universe at a sufficiently high-temperature. 
In order to end inflation the Higgs field has to tunnel to the other side of the potential barrier by nucleating bubbles~\cite{Coleman} 
that eventually collide and percolate.
Subsequently the Higgs field could roll down the potential, reheat the Universe and finally relax in the present true vacuum with $v=246$ GeV.
Whether the tunneling event happens depends on $H$ and $\Gamma$, the nucleation rate per unit time and volume.

If $ \Gamma \gg H^4 $, the Universe tunnels quickly in a few Hubble times, without providing sufficient inflation.
If $ \Gamma \ll H^4 $, the tunneling probability is so small that the process does not produce a sufficient number of bubbles inside a Hubble 
horizon that could percolate.
A graceful exit  would thus require $\Gamma/H^4$ to become larger than $1$ only after some time, 
but this is  impossible if both quantities are time-independent, as is the case for the pure SM embedded in standard gravity~\cite{Guth}.

A time-dependent $\Gamma/H^4$ necessarily requires the existence of an additional time-dependent order parameter. 
This can be realized in a scalar-tensor theory of gravity, where the value of the Planck mass is set by
a scalar field $\phi$, the Brans-Dicke scalar or dilaton. 
 This allows coupling  $\phi$ to the Ricci scalar $R$ via an interaction of the form
$f(\phi) R$,  where $f(\phi)>0$ thus sets the value of the Planck mass. The presence of such field makes
the Planck mass time-dependent, and therefore also $H$,  naturally leading to an increase in $\Gamma/H^4$.

This has been shown in early models~\cite{johri,extended,hyperextended} achieving power-law inflation,
which later turned out to be in tension with observations of the Cosmic Microwave Background (CMB), 
since it is difficult to get a nearly flat spectrum of perturbations~\cite{Liddle}. 
In refs.~\cite{hep-ph/0511207,astro-ph/0511396}, a stage of exponential expansion was naturally incorporated, 
later followed by a stage of power-law (even decelerated) expansion. 
In this way, it is possible to produce a flat nearly homogenous Universe during the exponential phase and, moreover, 
the quantum fluctuations in $\phi$ lead to the correct spectrum of perturbations. 
During the subsequent decelerated phase, $H$ decreases rapidly,  allowing the field trapped in the false minimum to tunnel through percolation of bubbles.
As discussed in \cite{Masina:2011aa}, after tunneling we require the field $\phi$ to relax to zero, which allows us to identify the present Planck mass
$M_{Pl}= 1/(8\pi G_N)^{1/2}$ $=1.22 \times 10^{19} /\sqrt{8 \pi} $ GeV with the Planck mass at inflation $M$ 
and, at the same time, to satisfy constraints from fifth-force experiments and time-dependence of the Newton constant $G_N$   \cite{Will:2005va}.

An alternative to scalar-tensor theories is given by models \cite{Masina:2012yd}
with a direct coupling of the Higgs field to an additional scalar field,
which induces a time-dependence directly into $\Gamma$ by flattening the barrier in the potential 
or it might be possible to achieve a graceful exit in other models 
 with an additional coupling of the Higgs field to $R$  \cite{Masina:2011aa}.

It is crucial to notice that  a graceful exit can be generically realized only if at the end of inflation there is a very shallow false minimum, 
otherwise the tunneling rate would be negligibly small,
since the probability is exponentially sensitive to the barrier \cite{Coleman}.
So, the shape of the potential is very close to the case in which there is just an inflection point. 
This leads to a powerful generic prediction\footnote{The relevant stage of inflation for predicting the amplitude of $r$ is the one at $50-60$
e-folds before the end. In models in which the potential is time-independent as the ones in \cite{Masina:2011aa} we necessarily have a shallow false minimum. 
Models with time-dependent $\Gamma$ can also be constructed, but even in those models it is
likely that $50$ e-folds before the end of inflation the potential well is generically not deep, since a too rapid variation of the potential in the
last stages of inflation would probably be in conflict with observations of the spectral index, as in \cite{Masina:2012yd}. }
for the scale of inflation and therefore for $r$. 
So, if the false vacuum is very shallow, the specific  model only affects the prediction for the spectral index of cosmological 
density perturbations $n_S$.
For instance, for a wide class of functions $f(\phi)$ the models considered in \cite{hep-ph/0511207} lead to $0.94 \lesssim n_S \lesssim 0.96$,
in agreement with the central value subsequently measured by WMAP \cite{Spergel:2003cb}.

\begin{figure}[t!]
 \includegraphics[width=6.7cm]{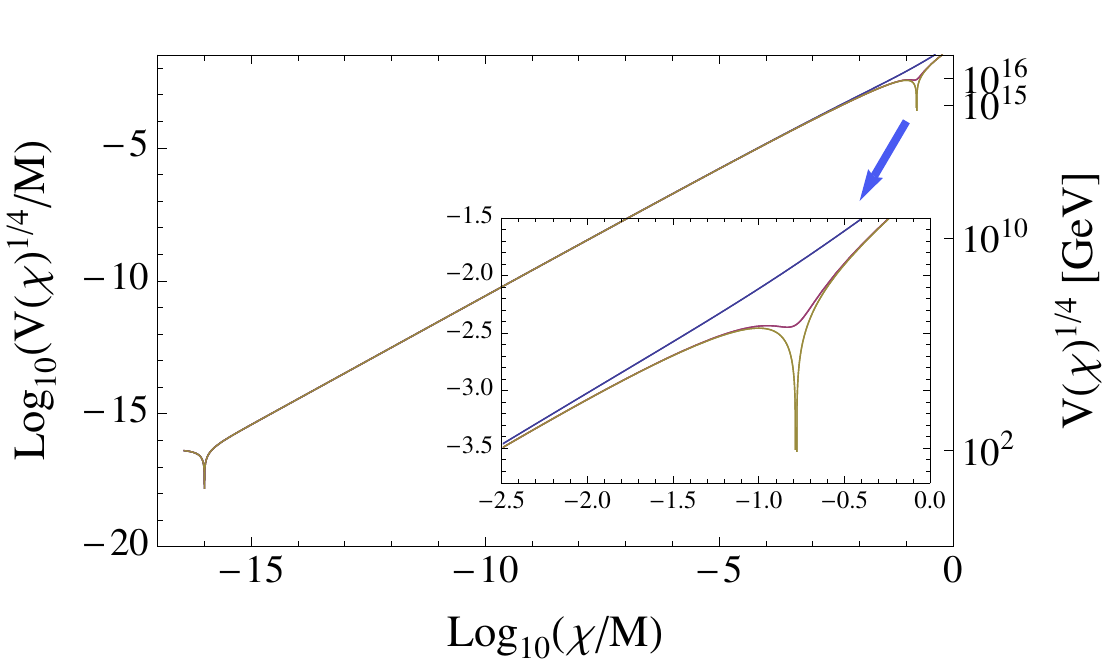}  
\caption{Higgs potential as a function of the Higgs field $\chi$. We fixed  $\alpha_3(m_Z)=0.1184$,
$m_t=171.8$ GeV and, from top to bottom, $m_H=125.2,125.158,125.157663$ GeV. 
In order to have a non-negligible tunneling probability $m_H$ should be determined with 15 significative digits, 
as we checked by numerically evaluating a bounce \cite{Coleman} solution.}
\label{fig-Vmin}
\vskip .2 cm
\end{figure}

Using 2-loop RGE and matching conditions, we studied the very specific values of the top and Higgs masses
allowing for the presence of a false minimum.
As an example, in fig.\ref{fig-Vmin} we display the Higgs potential for very specific values of $m_t$ and $m_H$. 
The extremely precise values shown in the caption are not to be taken sharply, because of a theoretical 
uncertainty of about $3$  GeV on $m_H$ and about $1$ GeV on $m_t$, which is intrinsic in the 2-loop RGE procedure \cite{hep-ph/0104016,strumia2}. 
As mentioned above and discussed in \cite{Masina:2011aa}, in order to have a sizable tunneling probability through the left side, the barrier must be very low, 
as is the case for the middle curve. 
For slightly smaller values of $m_H$ the second minimum becomes deeper and the tunneling probability essentially zero\footnote{For smaller 
values of $m_H$ the potential turns negative, so that 
the SM minimum at low energy becomes metastable \cite{hep-ph/0104016,strumia2}. }.

\begin{figure*}[t!]
\includegraphics[width=11.5cm]{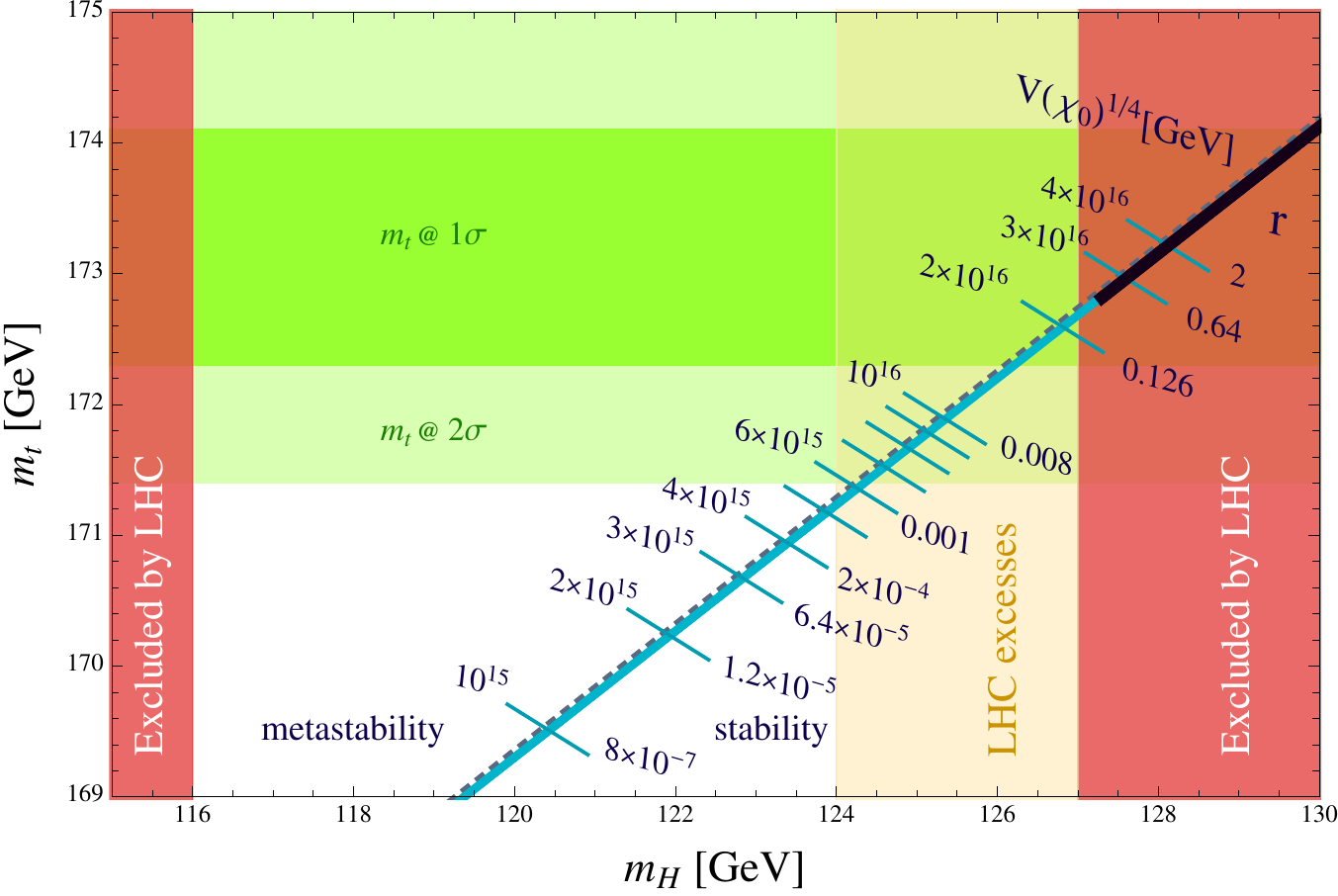}
\caption{ The solid line indicates the $m_t-m_H$ values compatible with a shallow Higgs false minimum, taking
$\alpha_s(m_z)=0.1184$ (central value of Particle Data Group  2011).  
The line has a (vertical) uncertainty  of $1$ GeV in $m_t$ and a (horizontal) one of $3$ GeV in $m_H$ due to the theoretical uncertainty of the 
2-loop RGE. The shaded horizontal bands are the $1\sigma$ and $2 \sigma$ ranges for $m_t=173.2\pm0.9$ GeV, 
according to the recent global SM electroweak precision fits \cite{GFitter}. 
Ticks along the solid line display the associated values of $V(\chi_0)^{1/4}$ in units of GeV and those of $r$. 
The black strip marks the values of $r$ already excluded \cite{Komatsu:2010fb}.}
\label{fig-mtmh}
\vskip .22 cm
\end{figure*}

Increasing (decreasing)  $m_t$, one has also to increase (decrease)  $m_H$ in order to develop the shallow false minimum; 
accordingly, the value of both $V(\chi_0)$ and $\chi_0$ increase (decrease).
The solid line in fig.\ref{fig-mtmh} shows the points in the $m_t-m_H$ plane where the shallow SM false minima exists.
We recall however that the line has a (vertical) uncertainty  of $1$ GeV in $m_t$ and a (horizontal) one of $3$ GeV in $m_H$.
The shallow false minima are just at the right of the dashed line marking the transition from stability to metastability.
Ticks along the solid line display the associated values of $V(\chi_0)^{1/4}$ in units of GeV.

The measurement of $m_t$ implies the lower bounds $m_H \gtrsim122.5$ GeV  and $V(\chi_0)^{1/4} \gtrsim 2.5 \times 10^{15} $ GeV, 
considering the theoretical errors intrinsic in the RGE.
On the other hand, as it is well-known and explained below, the scale of inflation cannot be too high,  $V(\chi_0)^{1/4}  \lesssim 2.5 \times10^{16}$ GeV, 
which leads to the constraint $m_H\lesssim 130$ GeV.
Remarkably, the allowed band for SM false vacuum inflation includes the region $124-127$ GeV, 
where ATLAS and CMS  \cite{HCP11} recorded excesses of events in the di-photon as well as 4-leptons channels. 
Moreover, now CMS has set the upper bound  $m_H\le 127$ GeV at $95\%$ CL, which further restricts the allowed region.

This striking coincidence of values deserves further exploration both experimentally,
by reducing the error on $m_t$ and $m_H$, and theoretically by improving the RGE\footnote{Note that additional particles 
could exist at high scales, modifying the running of $\lambda$.  If unification of gauge couplings is realized in nature,
the modification is however probably constrained to be small.}. 

The dominant source of the uncertainty in the RGE at present arises from the matching of the quartic Higgs coupling, known only at 1-loop. 
By varying the matching scale from about $125$ GeV (close to $m_H$) and about $175$ GeV (close to $m_t$) one finds that the value of $m_H$ 
leading to a shallow false minimum at the GUT scale changes by $1$ GeV.
Clearly the range of where to vary the matching scale is somewhat arbitrary, and in the literature one can find different choices. 
As in many other papers \cite{hep-ph/0104016,strumia2} using the same accuracy for the RGE as ours, we considered it
conservative to assign an error of $3$ GeV on $m_H$,  and an error of $1$ GeV on $m_t$.
This might overestimate the theoretical error, but in order to better understand it, one would need to know the 2-loop correction to the matching 
of the quartic Higgs coupling. Reasonably, one could expect that in this way the theoretical 
error on $m_H$ could be reduced down to $1$ GeV, which is comparable to the 
experimental precision on $m_H$ foreseen at LHC.
By varying $m_H$ within a range of $1$ GeV, one obtains in our scenario a
prediction for $m_t$ within a range of about $0.5$ GeV, call it $m_t^{pr}$. 
The experimental precision on $m_t$ needed to falsify our scenario depends 
on the difference between $m_t^{pr}$ and the experimental central value 
of $m_t$. The smaller this difference is, the more precision is needed 
on $m_t$.

A complementary way of testing the possibility that inflation started from the SM false vacuum at high energy
is to look at the tensor-to-scalar ratio of cosmological perturbations. 
The amplitude of  density fluctuations in the observed Universe as seen by the CMB and Large-Scale structure data  
is parametrized by the power spectrum in $k$-space
\be
P_s(k)=\Delta_R^2 \left( \frac{k}{k_0} \right)^{n_S-1} \,\,,
\ee
where $\Delta_R^2$ is the amplitude at some pivot point $k_0$.
We consider the best-fit value from \cite{Komatsu:2010fb}, $\Delta_R^2= (2.43 \pm 0.11)\times  10^{-9}$ 
 at $k_0=0.002 \,{\rm Mpc}^{-1}$.

In any inflationary model that can be analyzed through the slow-roll approximation \cite{Komatsu:2010fb}, there is a  relationship between the scale of inflation, 
the amplitude of density perturbations, and the amount of gravity waves that can be produced:
\be 
\Delta_R^2 =  \frac{2}{3 \pi^2}  \frac{1}{r} \frac{ V(\chi_0) }{ M^4  }  \,\,.
\label{eq-r}
\ee

If inflation actually started from a SM shallow false minimum, then each point in the $m_t-m_H$ plane has to
be associated with a specific value of $r$, as shown in fig.\ref{fig-mtmh}  via the lower row of ticks.
The upper limit on $r$ is at present about $0.2$ \cite{Komatsu:2010fb},  
which gives rise to the above mentioned upper bound $V(\chi_0)^{1/4} \lesssim 2.5 \times10^{16}$ GeV.
The lower bound on $m_t$ instead implies $r \gtrsim 10^{-4}$, partially at hand of future experimental sensitivity for various experiments
such as Planck \cite{Planck}, EPIC \cite{Bock:2009xw} and COrE \cite{Collaboration:2011ck}. 
Improving the top quark mass measurement and/or discovering the Higgs mass close to $126$ GeV could further constrain $r$ from below. 

The relationship between $r$, $m_t$ and $m_H$ is completely general in any
model with a SM shallow false vacuum. As we have mentioned, it is conceivable
that models could be constructed in which the false vacuum initially is not
shallow, but somehow the Higgs potential becomes time-dependent and is
lifted up,
making $\Gamma$ large and leading inflation to an end. In such models
only at the end of inflation the final shape of the potential is given
by the SM Higgs potential with a shallow minimum. 
In this case we can make two important statements.
First,  the height of the minimum during the observationally relevant
stage of inflation ({\it i.e.} $50-60$ e-folds before the end) is always at
most as high as the one in the shallow case: as a consequence, our
prediction on $r$ always applies strictly
as an upper bound for any model which uses the Higgs false minimum to
source inflation. So, if future experiments will measure $m_t$ and
$m_H$
accurately and if the RGE theoretical error is reduced, measuring $r$ above
the corresponding value displayed in fig.\,\ref{fig-mtmh} would rule out all such models.
Second, even in the case of a time-dependent Higgs potential, it is very
likely that at $50-60$ e-folds before the end of inflation the barrier is still
close to the shallow case, so fig.\,\ref{fig-mtmh} would probably apply even
in such models.

Summarizing, we argue that precision measurements of $r$, $m_t$, $m_H$, together with theoretical improvements of the SM RGE,
will represent a test of the hypothesis \cite{Masina:2011aa} that inflation occurred in the SM false vacuum at about $10^{16}$ GeV.

\section*{Acknowledgements}
We thank T. Hambye, A. Strumia and G. Villadoro for useful discussions.


\end{document}